\documentclass[twoside,12pt]{article}
\usepackage{amssymb, amsmath, amsfonts}

\begin{document}

\title{On a transformation between hierarchies of integrable equations}
\author{ Metin G{\" u}rses$^\dag$  and Kostyantyn Zheltukhin$^\ddag$ \\
{\small $^\dag$Department of Mathematics, Faculty of Sciences}\\
{\small Bilkent University, 06800 Ankara, Turkey}\\
{\small e-mail gurses@fen.bilkent.edu.tr}\\
{\small $^\ddag$Department of Mathematics, Faculty of Sciences}\\
{\small Middle East Technical University 06531 Ankara, Turkey}\\
{\small e-mail zheltukh@metu.edu.tr}\\}

\begin{titlepage}

\maketitle

\begin{abstract}
A transforation between a hierarchy of integrable equations
arising from the standard $R$-matrix construction on the algebra
of differential operators and  a hierarchy of integrable equations
arising from a deformation of
 the standard $R$-matrix is given.
\end{abstract}

{\it PASC: } 02.30.Ik


\end{titlepage}

In a recent paper \cite{SBl} a new hierarchy of integrable
equations has been constructed through the deformation of a
standard $R$-matrix on the algebra of pseudo-differential
operators. We give a transformation between the hierarchy
constructed in \cite{SBl} and a hierarchy obtained through a
standard $R$-matrix. The transformation is between corresponding
vector fields (i.e. symmetries).

 Let $\mathfrak {g}$ be the Lie algebra of pseudo-differential
operators
\begin{equation}
\mathfrak {g}=\left\{\sum\limits_{i\in {\bf Z}} u_i(x)D^i\right\}
\end{equation}
with the commutator $[L_1,L_2]=L_1L_2-L_2L_1$. The algebra $\mathfrak {g}$
 can be decomposed into Lie subalgebras
 ${\mathfrak {g}}_{\ge k}= \left \{ \sum\limits_{i \ge k} u_i(x)D^i \right\}$
 and , ${\mathfrak {g}}_{i< k}=\left\{\sum\limits_{i< k} u_i(x)D^i \right\}$
where $k=0,1,2$ (only for such  $k$ one has Lie subalgebras). The
standard $R$-matrix  is given by $R_k=\frac{1}{2}(P_{\ge
k}-P_{<k})$, where $P_{\ge k}$ and $P_{<k}$ are projection
operators on ${\mathfrak {g}}_{\ge k}$ and ${\mathfrak {g}}_{<k}$,
respectively. The  Lax hierarchy is
\begin{equation}\label{hierarchy}
L_{t_n}=[R(L^n),L]=[(L^n)_{\ge k},L], \qquad L\in {\mathfrak {g}},
\qquad n=1,2\dots\, .
\end{equation}
 The above equations involves infinitely many fields.
To have a consistent closed equations with a finite number of
fields we restrict the Lax operators as  follows
\begin{equation}\label{lax1}
k=0\quad L_0=D^N+u_{N-2}D^{N-2}+\dots+u_1D+u_0
\end{equation}
\begin{equation}\label{lax2}
k=1\quad L_1=D^N+u_{N-1}D^{N-1}+\dots+u_0+D^{-1}u_{-1}
\end{equation}
\begin{equation}\label{lax3}
k=2\quad L_2=u_ND^N+u_{N-1}D^{N-1}+\dots+D^{-1}u_{-1}.
+D^{-2}u_{-2}
\end{equation}
See \cite{Bl} for more details on the $R$-matrix formalism.

Recently  in \cite{SBl}  the deformations of the above
$R$-matrices were introduced. Most of the introduced deformed
$R$-matrices do not lead to the new hierarchies.  A new hierarchy
is obtained through a deformation of $R$-matrix
$R_1=\frac{1}{2}(P_{\ge 1}-P_{<1})$. Let $P_{=i}(L)=(L)_{=i}$
denotes coefficient of $D^i$ in the expansion of $L\in \mathfrak
{g}$. Then the deformed $R$-matrix is
\begin{equation}
\tilde R=\frac{1}{2}(P_{\ge 1}-P_{<1})+\varepsilon P_{=0}(\cdot)D,
\end{equation}
where $\varepsilon$ is a deformation parameter. The hierarchy is
\begin{equation}
 L_{t_n}=[\tilde R( L^n),L],  \qquad L\in {\mathfrak {g}}, \qquad n=1,2\dots\, .
\end{equation}
The above equations involves infinitely many fields, to have a
consistent  closed equation with finite number of fields we
restrict the Lax operator as $\tilde
L=u_ND^N+u_{N-1}D^{N-1}+\dots+u_0+D^{-1} u_{-1}$. Then the new
hierarchy is
\begin{equation}  \label{hierarchynew}
\tilde L_{t_n}=[(\tilde L^n)_{\ge 1}+\epsilon (\tilde
L^n)_{=0}D,\tilde L], \qquad n=1,2 \dots,
\end{equation}
 note that $\tilde L=L_2|_{u_{-2}=0}$. See \cite{SBl} for more details.

In this work we shall show that  the new hierarchy
(\ref{hierarchynew})   is related to the hierarchy corresponding
to $R$-matrix $R_2$ with reduced Lax operator $\tilde L=
L_2|_{u_{-2}=0}$. So we relate hierarchy (\ref{hierarchynew}) to
the hierarchy
\begin{equation}\label{hierarchyred}
\tilde L_{t_n}=[(\tilde L^n)_{\ge 2},\tilde L], \qquad
n=1,2\dots\, .
\end{equation}
We note that both hierarchies have the same Lax operator.

The  construction of the transformation is based on  expressing
$(\tilde L^n)_{=1}$ and $(\tilde L^n)_{=0}$  in terms of
coefficients of $[(\tilde L^n)_{\ge 2},\tilde L]$, for $ n \in
{\bf N}$.

\vspace{0.3cm}

 \noindent {\bf Proposition 1}.\,{\it Let
$\tilde{L}=L_{2}|_{u_{-2}=0}$, then
\begin{eqnarray}
([(\tilde L^n)_{\ge 2},\tilde L])_{=N}=-([(\tilde
L^n)_{=1}D,\tilde L])_{=N},\label{eqlL1} \\
 ([(\tilde L^n)_{\ge 1},\tilde L])_{=N-1}=-([(\tilde L^n)_{=0},\tilde L])_{=N-1}. \label{eqlL0}
\end{eqnarray}
for all $N$ ($N$ is order of operator $\tilde L$).}

\vspace{0.3cm}

 \noindent {\bf Proof}.\, Comparing powers of $D$ on
the right  and left hand side of the equality
\begin{equation}
[(\tilde L^n)_{\ge 1},\tilde L]=-[(\tilde L^n)_{< 1},\tilde L],
\end{equation}
 we have
\begin{equation}
([(\tilde L^n)_{\ge 1},\tilde L])_{=N}=0.
\end{equation}
 Then
\begin{equation}
([(\tilde L^n)_{\ge 2},\tilde L])_{=N}=-([(\tilde
L^n)_{=1}D,\tilde L])_{=N}.
\end{equation}
In the same way, comparing powers of $D$ on the right  and left
hand side of the equality
\begin{equation}
[(\tilde L^n)_{\ge 0},\tilde L]=-[(\tilde L^n)_{< 0},\tilde L]
\end{equation}
 we have
\begin{equation}
([(\tilde L^n)_{\ge 0},\tilde L])_{=N-1}=0.
\end{equation}
So,
\begin{equation}
([(\tilde L^n)_{\ge 1},\tilde L])_{=N-1}=-([(\tilde
L^n)_{=0},\tilde L])_{=N-1}.
\end{equation}

\vspace{0.3cm}

\noindent The above equalities (\ref{eqlL1}) and (\ref{eqlL0})
allows us to express  $(\tilde L^n)_{=1}$ and $(\tilde L^n)_{=0}$
in terms of coefficients of~$[(\tilde L^n)_{\ge 2},\tilde L]$ for
all $N$. Let us give an example for $N=1$.

\vspace{0.3cm}

\noindent {\bf Proposition 2}.\,{\it
 Consider the Lax   operator  $\tilde{L}=uD+v+D^{-1}w$. Let
\begin{equation}
[(\tilde L^n)_{\ge 2},\tilde L]=f_nD+g_n+D^{-1}h_n,
\end{equation}
which gives the hierarchy (\ref{hierarchyred}) with the standard
$R$-matrix and
\begin{equation}
[(\tilde L^n)_{\ge 1}+(\tilde L^n)_{=0}D,\tilde
L]=p_nD+q_n+D^{-1}r_n,
\end{equation}
which gives the hierarchy (\ref{hierarchynew}) with the deformed
$R$-matrix, $n=1,2\dots\,$. The coefficients $f_n,\, g_n,\, h_n,
p_n,\, q_n,\, r_n$ are functions of $u,\, v,\, w$ and their
derivatives. Then
\[
(p_n,\,q_n,\,r_n)^T={\cal T}\,(f_n,\,g_n,\,h_n)^T
\]
where
\begin{equation}\label{opr1}
{\cal T}=\left (
\begin{array}{lll}
\varepsilon u_xD^{-1} v_xD^{-1} u^{-2} - \varepsilon uv_xD^{-1}
u^{-2} &
\varepsilon u_xD^{-1} u^{-1} -\varepsilon  & 0\\
& &\\
 uv_xD^{-1} u^{-2}  +  \varepsilon v_xD^{-1}v_xD^{-1} u^{-2} &
1+ \varepsilon v_xD^{-1}u^{-1} & 0\\
& & \\
 ( (uw)_x+\varepsilon wv_x)D^{-1} u^{-2} &
\varepsilon wu^{-1} +  \varepsilon w_xD^{-1}u^{-1}& 1 \\
+ \varepsilon w_xD^{-1} v_xD^{-1} u^{-2} +wu^{-1}& & \\
\end{array} \right )
\end{equation}
}

 \vspace{0.3cm}

\noindent {\bf Proof.}\, Let $(\tilde L^n)_{=1}=A_n$ and $(\tilde
L^n)_{=0}=B_n$. The equality (\ref{eqlL1}) implies that
\begin{equation}
f_n=-([A_nD,\tilde L])_{=1},0
\end{equation}
hence, we can find
\begin{equation}\label{An}
A_n= uD^{-1}u^{-2}f_n.
\end{equation}
 Using the equality
(\ref{eqlL0}) we have
\begin{equation}
g_n + ([A_nD,\tilde L])_{=0} =-([B_n,\tilde L])_{=0},
\end{equation}
hence, we can find
\begin{equation}\label{Bn}
B_n=D^{-1}(u^{-1}g_n+v_xD^{-1}u^{-2}f_n).
\end{equation}
From the equality
\begin{equation}
[(\tilde L^n)_{\ge 1}+\varepsilon (\tilde L^n)_{=0}D,\tilde L]=
[(\tilde L^n)_{\ge 2},\tilde L]+ [(A_n+\varepsilon B_n)D,\tilde L]
\end{equation}
 we can find the transformation between the vector fields
\begin{equation}
\begin{array}{lll}
p_n&=&u_x\varepsilon B_n-u\varepsilon B_{n,x}\\
q_n&=&g_n+v_x(A_n+\varepsilon B_n)\\
r_n&=&h_n+(w(A_n+\varepsilon B_n))_x\\
\end{array}
\end{equation}
 where  $A_n$ and $B_n$ are given by (\ref{An}) and (\ref{Bn})
respectively. Thus we obtain the transformation operator ${\cal
T}$ in (\ref{opr1}).

 \vspace{0.3cm}

\noindent If we apply operator  ${\cal T}$
 to the simple symmetry $(u_{x},v_{x},w_{x})^T$ we
obtain $(0,0,0)^T$. Applying the operator ${\cal T}$ to
$(0,0,0)^T$ we get
\begin{equation} \left (
\begin{array}{l}
p_{1}\\
q_{1}\\
r_{1}
\end{array}
\right )=\left (
\begin{array}{l}
\varepsilon (vu_{x}-uv_{x}+u_{x})\\
uv_{x}+\varepsilon (vv_{x}+v_{x})\\
(uw)_{x}+\varepsilon (vw)_{x}+\varepsilon w_{x}
\end{array}
\right ).
\end{equation}
This is the deformed system (\ref{hierarchynew}) for $n=1$ (with
the inclusion of the symmetry $(u_{x},v_{x},w_{x})^{T}$),
\cite{SBl}. If we take symmetry of the hierarchy
(\ref{hierarchyred}) corresponding to $n=2$ (this is the reduced
system \cite{Bl}, \cite{Bl1})
\begin{equation}
\left (
\begin{array}{l}
f_{2}\\
g_{2}\\
h_{2}\\
\end{array}
\right )= \left (
\begin{array}{l}
u^2u_{xx} + 2u^2v_x\\
u^2v_{xx} + 2u(uw)_x\\
-(u^2w)_{xx}
\end{array}
\right )
\end{equation}
and apply the operator $\cal T$ to this symmetry we obtain a
second symmetry of the hierarchy (\ref{hierarchynew})
\begin{equation}
 \left (
\begin{array}{l}
p_{2}\\
q_{2}\\
r_{2}
\end{array}
\right )= \left (
\begin{array}{l}
\varepsilon u_xv^2 -2\varepsilon uvv_x -2\varepsilon u^2w_x-\varepsilon u^2v_{xx}\\
\\
2uu_xw+2uvv_x+2u^2w_x+uu_xv_x+u^2v_{xx}+\varepsilon v^2v_x
+2\varepsilon uv_xw
+\varepsilon uv_x^2\\
\\
 2u_xvw+2uv_xw +2uvw_x -u_x^2w-3uu_xw_x-uu_{xx}w-u^2w_{xx}+\\
2\varepsilon u_xw^2+2\varepsilon vv_xw+\varepsilon u_xv_xw+
\varepsilon v^2w_x+4\varepsilon uww_x+\varepsilon uv_xw_x+
\varepsilon uv_{xx}w\\
\end{array}
\right ).
\end{equation}

\vspace{0.3cm}

\noindent {\bf Remark.}\, In the example above we have constructed
the transformation ${\cal T}$ for hierarchies with Lax operator of
order one. In the same way we can construct the transformation
between hierarchies with Lax operator of any order $N$. The
operator ${\cal T}$ is not a recursion operator. It maps the
symmetries of one system  of  evolution equations to symmetries of
another system of  evolution equations.

 \vspace{0.5cm}
This work is partially supported by the Turkish Academy of
Sciences and by the Scientific and Technical Research Council of
Turkey

\end{document}